\documentclass[longbibliography,noeprint,noshowpacs,nopreprintnumbers,twocolumn,pra,showpacs,superscriptaddress,amsmath,amssymb,citeautoscript,aps,10pt]{revtex4-2}
\usepackage{graphicx}
\usepackage{dcolumn}
\usepackage{color}
\usepackage{bm}
\usepackage{chngcntr}
\usepackage[hidelinks]{hyperref}
\hypersetup{
    colorlinks,
    citecolor=blue,
    filecolor=blue,
    linkcolor=blue,
    urlcolor=blue
}

\graphicspath{{Figures/}{./Figures/}{.}}

\allowdisplaybreaks

\begin{document}

\title{Dynamical Scarring from Scrambling in Two Dimensional Topological Materials}
\author{Dominik Szpara}
\affiliation{Institute of Physics, Maria Curie-Sk\l{}odowska University, 20-031 Lublin, Poland}
\author{Szczepan Głodzik}
\affiliation{Institute of Physics, Maria Curie-Sk\l{}odowska University, 20-031 Lublin, Poland}
\author{Nicholas Sedlmayr}
\email[e-mail: ]{sedlmayr@umcs.pl}
\affiliation{Institute of Physics, Maria Curie-Sk\l{}odowska University, 20-031 Lublin, Poland}

\date{\today}

\begin{abstract}
Out-of-time ordered correlators are a probe of how the information of an initial perturbation is effectively scrambled under unitary time evolution, widely used to study quantum chaos. They have also been used to demonstrate that information is trapped in the zero dimensional edge modes of topological insulators and superconductors, and does not become scrambled. Here we study scrambling in two dimensional topological models. In the bulk the butterfly velocity, the speed at which the out-of-time ordered correlator spreads, gains a directional dependence from the underlying lattice. Furthermore when there are chiral or helical edge modes present these cause a form of dynamical scarring. The information about an initial perturbation on the boundary of the system travels around the edge, carried by the edge modes, but is not scrambled over very long time scales. The direction and speed of the scars are given by the velocities of the linearly dispersing edge modes. We further show that these scars do not interact, passing through each other. We back up these results with analytical and numerical calculations on exemplary models. 
\end{abstract}

\maketitle

\section{Introduction}

It is now well understood that phases of matter can be characterized not only by symmetry properties and order parameters, but also by topological invariants~\cite{Hasan2010,Chiu2016,Sato2017,Rachel2018,Xie2021,Yang2024a}. For example for the gapped band structures of insulators and mean field superconductors a topological index can be defined\cite{Hasan2010,Chiu2016}. This index distinguishes different topological phases, and the interest in them is in no small part due to the bulk-boundary correspondence: the bulk invariant predicts the existence of topologically protected modes at the boundaries and interfaces of the materials. Direct measurement of the index is often not straightforward, and though boundary modes can be spectroscopically probed their topological protection is often unclear. Therefore tools which aid the investigation of the boundary modes or topology are highly sought after both experimentally and theoretically.

The method we focus on here is based on out-of-time ordered correlators (OTOCs). OTOCs probe how correlations and perturbations are scrambled during time evolution. They were originally introduced to study superconductivity~\cite{Larkin1969}, though now are more likely to be applied to studies of chaos in quantum systems~\cite{Chowdhury2017,Xu2020b} as the OTOC gives a measure of how information becomes spread out, and practically speaking unrecoverable, in a quantum system despite the unitary dynamics.

As the experimental control of quantum matter has steadily increased it has become possible to study the dynamics of quantum information~\cite{Kinoshita2006,Cheneau2012,Landsman2019}, and it is also possible to measure OTOCs themselves experimentally~\cite{Swingle2016,Garttner2017,Li2017,Lewis-Swan2019,Nie2020,Blok2021,Sundar2022,Weinstein2022}. This has sharpened the question of how information spreads and is effectively destroyed during the dynamics of many-body systems. Classically it is understood that a marker of chaos is the exponential divergence in phase space of states which began close together. This is bound by the Lyapunov exponent and a quantum Lyapunov exponent $\lambda_L$ has been found to bound the growth of information scrambling when characterized by an OTOC~\cite{Maldacena2016}. Local information spreads through the system at a so-called butterfly velocity, $v_b$, a process which is referred to as the scrambling of the information. Once the information is spread out, it can not be easily reconstructed, even though this process is of course unitary and technically speaking reversible. This can also be thought of as an effective Lieb-Robinson bound for the particular states under investigation~\cite{Lieb1972,Roberts2016}. Scrambling in quantum systems has therefore sometimes been regarded as the quantum analogue of chaos in a classical system~\cite{Chowdhury2017,Xu2020b}.

Depending on the nature of the model being investigated scrambling can be exponentially slow~\cite{McGinley2019b,Okvatovity2019} or fast~\cite{Belyansky2020}. OTOCs and scrambling have also been used to investigate the thermalization of a quantum system~\cite{Deutsch1991,Srednicki1994,Rigol2007,Sirker2014a,Deutsch2018}, relating it to entanglement growth during the dynamics~\cite{VonKeyserlingk2018,Alba2019,Modak2020}. Other applications have been studying localization~\cite{Swingle2017,Slagle2017}, and investigating a variety of models including the Sachdev-Ye-Kitaev model~\cite{Maldacena2016a}, the $O(N)$ model~\cite{Chowdhury2017}, hard-core bosons~\cite{Patel2017, Lin2018a} and two dimensional Bose-Hubbard lattices~\cite{Tripathy2024}, spin chains~\cite{Riddell2019}, topological insulators~\cite{Sedlmayr2023}, and Floquet systems~\cite{Zamani2022}. OTOCs have also been useful in characterizing ergodic-nonergodic transitions~\cite{Buijsman2017}, equilibrium phase transitions~\cite{Dag2019}, topological phase transitions~\cite{Bin2023}, excited-state quantum phase transitions~\cite{Wang2019d}, and dynamical quantum phase transitions~\cite{Heyl2018,Heyl2018a}.

With regard to topological systems it has been demonstrated that at the edges of one dimensional topological models the correlations measured by OTOCs become stuck in the edge modes~\cite{Sedlmayr2023}, remaining stationary for long time scales and not becoming scrambled. This could therefore be thought of as a form of scarring~\cite{Turner2018,Mondal2020}, \emph{i.e.}~information about the initial state survives during the dynamics. In this case  scarring refers to the behavior of the \emph{OTOC}. In the bulk the usual scrambling is observed, with no discernible role played by the topology. In two dimensional topological models this raises the natural question of what happens when chiral or helical modes are present. Is there a form of mobile scarring with the correlations moving around the edge, or does the OTOC in any case show scrambling on the boundary. It is this question we turn to in this article. We consider scrambling caused both by the bulk and boundary modes, finding that information in the boundary modes is long lived but dynamical, moving in the direction of the chiral or helical edge modes. This dynamical scarring of the OTOC is shown to be uninfluenced by disordered edges as may be expected for a phenomena of topological origin. We nonetheless want to stress that this is a consequence of a lack of operator spreading during time evolution, and not simply to ballistic transport of particles in edge modes.

This article is structured as follows. In section \ref{sec_otoc} we introduce the definition of the OTOCs and the method we use to calculate them in our finite open models. In section \ref{sec_mod} we introduce the exemplary topological models we will study. In section \ref{sec_analytics} we calculate analytically the scrambling, including the scrambling for an effective one dimensional model of the topologically protected edge modes. Section~\ref{bulk} investigates scrambling in the bulk, and sections \ref{chiral} and \ref{helical} demonstrate the scarring on the edges of chiral and helical models. In section \ref{conclusion} we conclude and discuss our results.

\section{Out-of-Time Ordered Correlators}\label{sec_otoc}

To investigate whether and how information becomes scrambled in our topological model, we use the out-of-time ordered commutator and its associated out-of-time ordered correlator. For local unitary perturbations $\hat V_j$ and $\hat W_j$ we have for the OTOC~\cite{Swingle2016,Roberts2016}
\begin{equation}\label{c1}
	C_{j,j_0}(t)=\left\langle\left[\hat W_{j_0}(t),\hat V_j\right]^\dagger\left[\hat W_{j_0}(t),\hat V_j\right]\right\rangle\,,
\end{equation}
where $\hat W_j(t)=e^{i \hat Ht}\hat W_je^{-i \hat Ht}$ is the time evolution of $\hat W_{j_0}$. In this form the OTOC is explicitly Hermitian, though we note alternative non-Hermitian definitions have also been used~\cite{Larkin1969,Dora2017a,Heyl2018}. It is convenient to introduce a related correlator defined by
\begin{equation}
	F_{j,j_0}(t)=\left\langle \hat W_{j_0}^\dagger(t)\hat V_j^\dagger \hat W_{j_0}(t)\hat V_j\right\rangle
\end{equation}
where $C_{j,j_0}(t)=2(1-\Re[F_{j,j_0}(t)])$. As we are interested in the behavior of topological models, and in particular the role their edge modes play in scrambling, the quantum average is taken over the ground state of $\hat H$: $|\psi_0\rangle$. Elsewhere the role of quenches in scrambling has been investigated, in this case the time evolving Hamiltonian $\hat H$ is different to the Hamiltonian of the ground state. However this appeared to play no role~\cite{Sedlmayr2023} and therefore we focus here on the case where $|\psi_0\rangle$ is the ground state of the time evolving Hamiltonian. We consider different Hamiltonians $\hat H$ with different topological symmetry classes, different topological phases, and different lattices. We can also consider composites of different Hamiltonians.

To calculate the out-of-time ordered correlator for a general free fermion system \emph{without} translational invariance we turn to the correlation matrix
\begin{equation}
{\bm M}=\langle\psi_0|\hat\Phi^\dagger\hat\Phi|\psi_0\rangle\,.
\end{equation}
$\hat\Phi$ is the appropriate single particle annihilation operator written in some basis: $\hat\Phi=(\hat c_1,\hat c_2,\ldots)$. Using a similar method to that used for the Loschmidt amplitude~\cite{Levitov1996,Klich2003,Rossini2007,Sedlmayr2018} one can find the following expression~\cite{Sedlmayr2023}:
\begin{equation}\label{eq:f}
	F_{j,j_0}(t)=\det\left[1+ {\bm M}\left( {\bm W}_{j_0}^\dagger(t){\bm V}_j^\dagger {\bm W}_{j_0}(t){\bm V}_j-1\right)\right]\,.
\end{equation}
Here all matrices are naturally written in the same basis. This provides a convenient and fast method for calculating the OTOC using exact diagonalization of the Hamiltonian.

For an integrable system in one dimension the OTOC is expected to follow an exponential increase given by~\cite{Lin2018,Xu2020a,Xu2022}
\begin{equation}\label{cfit}
    C_{j,j_0}(t)\sim e^{\lambda_L\frac{\left(t-\frac{a|j-j_0|}{v_b}\right)^{3/2}}{t^{1/2}}}
\end{equation}
where $\lambda_L$ is the Lyapunov exponent and $v_b$ is the butterfly velocity. For two dimensions such a simple expression is not generally possible, as we will show. In general the butterfly velocity and Lyapunov exponents depend on the direction from the perturbation one is considering, with the lattice geometry playing an important role.

As we are interested in scrambling in two dimensions the lattice indices $j$ and $j_0$ enumerate a two dimensional lattice. It is convenient to define $\vec{r}=\vec{r}_j$ and $\vec{r}_0=\vec{r}_{j_0}$ as the real space lattice sites labeled by $j$ and $j_0$ respectively. Our lattice spacing between nearest neighbor sites is $a$ and we set $a=1$ along with $\hbar=1$ throughout, retaining it only where the dimensions are useful to note.

\section{Models}\label{sec_mod}

We will consider two representative models to demonstrate scrambling, or its absence, in two dimensional topological materials. We take a paradigmatic Chern model on a square lattice, and the $\mathbb{Z}_2$ Kane-Mele model on a honeycomb lattice. These two models allow us to cover different kinds of boundary modes and different Bravais lattices, covering both chiral and helical boundary modes.

The first model is a two dimensional generalization of the spinless Kitaev chain to a $p+ip$ topological superconductor, which we will refer to as the Kitaev lattice~\cite{Sedlmayr2015b}. This model has chiral edge modes in the topologically non-trivial phases moving either clockwise or counter-clockwise depending on the sign of the Chern number. The Hamiltonian is given by
\begin{equation}\label{kit_ham}
\hat{H}=-\sum_{j,\ell}\Psi^\dagger_{j}\mu{\bm\tau}^z\Psi_{j}
-\sum_{\langle j,\ell\rangle}\Psi^\dagger_{j}\left[J{\bm\tau}^z-
i\Delta(\vec{d}_{j,\ell}\times\vec{{\bm\tau}})_z\right]\Psi_{\ell}\,.
\end{equation}
where $\Psi^\dagger_{j}=\{c^\dagger_{j},c_{j}\}$ with $c_{ j}^{(\dagger)}$ annihilating (creating) a spinless particle on a square lattice labeled by $j$ and $\vec{d}_{j,\ell}$ is the vector from site $j$ to site $l$. $J$ is the hopping strength and $\Delta$ the p-wave pairing strength with $\mu$ the chemical potential. We set $J=1$ throughout. This model has only a particle-hole symmetry and the invariant is the usual Chern number which can take values $\nu\in\{-1,0,1\}$. See for example \onlinecite{Maslowski2024b} for the explicit topological phase diagram.

For ease we take just several exemplary points in the phases with parameters \begin{align}
    (\Delta/J,\mu/J,\nu)\in&\{(0.45, 6, 0),(0.45, 2, 1),
    \nonumber\\&(0.45, -2, -1),(-0.45, 2, 1)\}.
\end{align}
Unless otherwise noted the $\nu=1$ phase will refer to the case where $\Delta>0$. We will also focus here on the perturbations
\begin{equation}\label{pert}
    \hat{V}_j=\hat{W}_j=e^{i\alpha\Psi_j^\dagger\tau^z\Psi_j},
\end{equation}
with $\alpha=5$ for the numerical calculations. We have checked perturbations along different orientations in the $\tau$ subspace and find the results do not qualitatively depend on this choice. The velocity of the edge state can be obtained from the bandstructure. For $|\Delta|=0.45J$ and $\mu=2J$ we find $v_F\approx 0.896 aJ$ where $a$ is the lattice spacing and $\hbar=1$, as it will be throughout this article.

The second model we use is the Kane-Mele model~\cite{Kane2005} with the staggered potential and Rashba coupling set to zero, which is a $\mathbb{Z}_2$ topological insulator with helical edge modes in the topologically non-trivial phase. The model is
\begin{align}\label{km_ham}
    \hat H=&-t\sum_{\langle j,\ell\rangle}\Phi^\dagger_{j}\Phi_\ell
    +
    i\lambda_S\sum_{\langle\langle j,\ell\rangle\rangle}\nu_{j\ell}\Phi^\dagger_{j}\tau^z\Phi_\ell\,.
\end{align}
$\langle j,\ell\rangle$ and $\langle\langle j,\ell\rangle\rangle$ denote nearest neighbor and next-nearest neighbor hopping respectively on a honeycomb lattice. In this case $\Phi^\dagger_{j,\ell}=\{c^\dagger_{j,\uparrow},c^\dagger_{j,\downarrow}\}$ with $c_{ j,\tau}^{(\dagger)}$ annihilating (creating) a particle with spin $\tau$ at site $j$. We then have hopping $t$, and $\lambda_S$ the mirror symmetric spin-orbit coupling. The factor $\nu_{j\ell}=\pm1$ depends on the direction of the hopping to the second neighbor. For this model we focus on the topologically non-trivial phase which hosts two counter-propagating helical edge modes and in the simulations we set $t=1$ and $\lambda_S=0.1t$.
The perturbation is
\begin{equation}\label{z2pert}    
\hat{V}_j=\hat{W}_j=e^{i\alpha\Phi_j^\dagger\tau^z\Phi_j},
\end{equation}
and again we set $\alpha=5$.

\begin{figure*}
    \centering
    \includegraphics[width=0.9\textwidth]{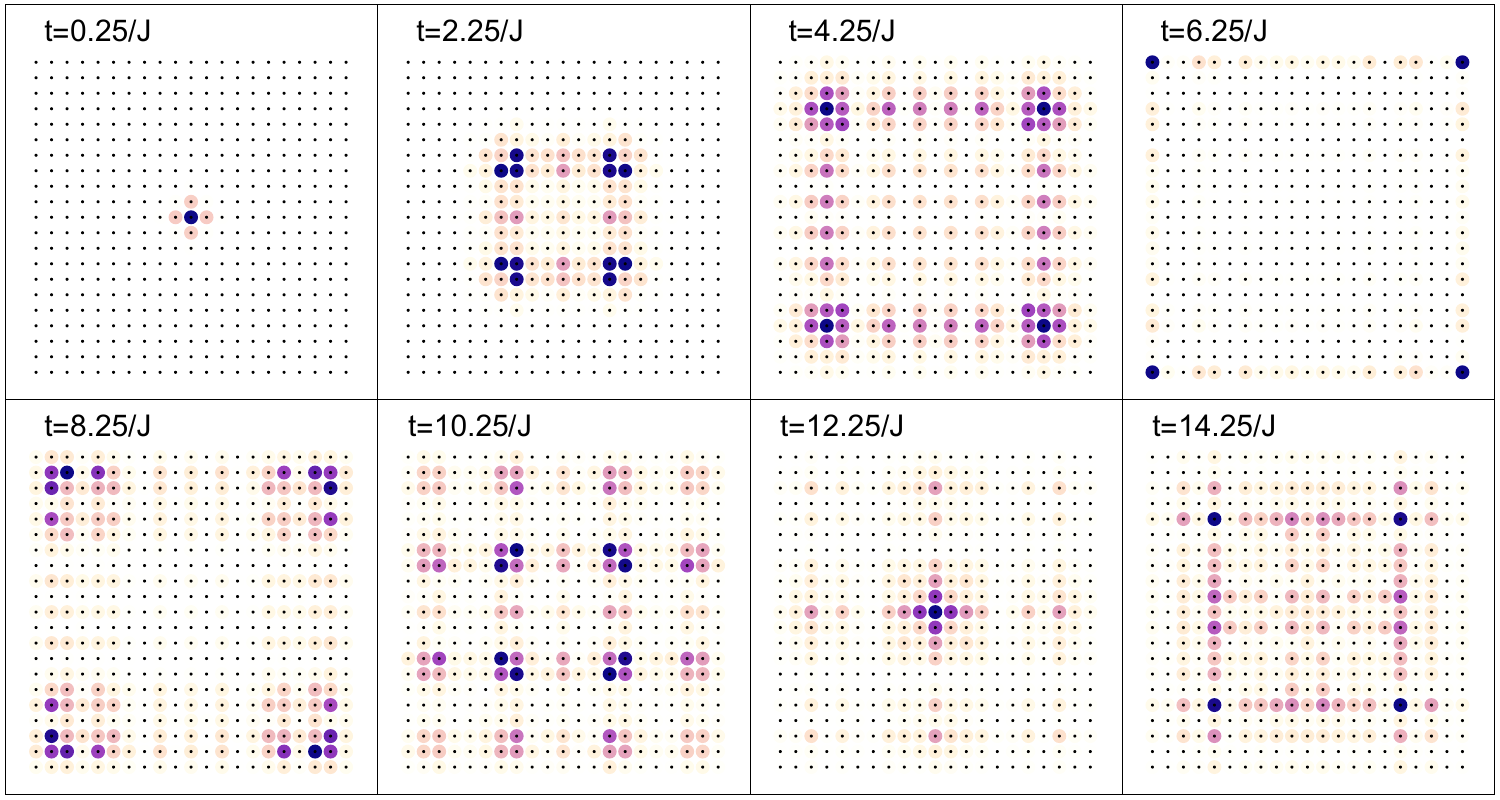}\\
    \includegraphics[width=0.35\textwidth]{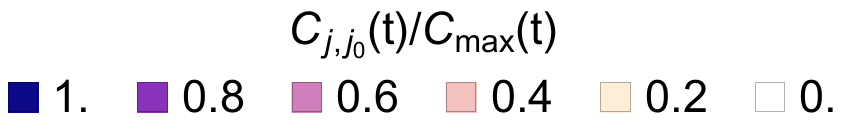}
    \caption{The OTOC $C_{j,j_0}(t)$ at different time steps following a perturbation in the center, at site $\vec{r}_0=(11,11)a$, in the topologically trivial phase with $\nu=0$, see main text for details. Each point is a lattice site. Note that the normalization of the color scheme is for each time step, $C_{\rm max}(t)=\textrm{max}[C_{j,j_0}(t)]$. Here we check the bulk scrambling which spreads throughout the system with a characteristic velocity,before scattering from the edges.}
    \label{fig:otocm0}
\end{figure*}

\section{Analytical Calculation}\label{sec_analytics}

For the non-interacting fermionic models under consideration it is possible to obtain some results analytically. Nonetheless, even in this case, not everything can be calculated analytically. For the finite size open boundary system where we have edge modes we must rely on numerics, see the next section. For the analytical calculations we will focus here on two main cases. A chiral edge mode along a translationally invariant edge, and the bulk scrambling for a gapped two band model.

We start with a general two band model, and will then specify whether we consider chiral linearly dispersing one dimensional modes or the topological superconductor, \eqref{kit_ham}, in the bulk. To model the chiral edge modes we take a continuum Hamiltonian with both left and right movers present and assume the left moving states are empty. We shall see this introduces an artifact into our calculation which we must neglect, but allows an efficient calculation of the behavior on the edge of our two dimensional model. Expanding the exponential of Eq.~\eqref{pert} and rearranging this can be rewritten as
\begin{equation}
    \hat{V}_j=1+\Psi^\dagger_j
        \underbrace{\begin{pmatrix}
         e^{i\alpha}-1&0\\0&e^{-i\alpha}-1
        \end{pmatrix}}_{\equiv A_\alpha}
    \Psi_j.
\end{equation}
For the OTOC we therefore require the commutator
\begin{equation}\label{c2}
	C_{j,j_0}(t)
    =\left\langle\left|\left[\Psi^\dagger_{j_0}(t)A_\alpha\Psi_{j_0}(t), \Psi^\dagger_{j}A_\alpha\Psi_{j}\right]\right|^2\right\rangle\,.
\end{equation}
To proceed we must transform to a more convenient basis: the eigenbasis of our Hamiltonian.

For convenience we will suppress the explicit dimension of the momentum and position, reinstating them only when necessary. The first step is then to make a standard Fourier transform. Following this we can rotate to the eigenbasis of $\Psi^\dagger_k\mathcal{H}_k\Psi_k=\tilde{\Psi}^\dagger_k\tilde{\mathcal{H}}_k\tilde{\Psi}_k$ such that $\tilde{\mathcal{H}}_k$ is diagonal and $\tilde{\Psi}_k=U^\dagger_k\Psi_k=(a_k,b_k)^T$. Let us assume that $\mathcal{H}_k=h^x_k\tau^x+h^y_k\tau^y$ is in the $x-y$ plane of its subspace. The rotation can be written as
\begin{equation}
    U_k=\frac{1}{\sqrt{2}}\begin{pmatrix}
        1&-1\\1&1
    \end{pmatrix}e^{i\frac{\theta_k}{2}\tau^z},
\end{equation}
with $\tan\theta_k=h^y_k/h^x_k$. We now find
\begin{equation}
    \Psi_j(t)=\frac{1}{\sqrt{N}}\sum_ke^{ikj}U_k
        \begin{pmatrix}
            e^{i\epsilon_kt}&0\\0&e^{-i\epsilon_kt}
        \end{pmatrix}
    \tilde{\Psi}_k,
\end{equation}
as the eigenenergies are $\pm\epsilon_k$. We therefore have
\begin{align}
    \Psi^\dagger_j(t)A_\alpha\Psi_j(t)=&\frac{1}{N}\sum_{k,q}e^{ikj-iqj}
    \tilde{\Psi}^\dagger_q
    M_{qk}(t)
    \tilde{\Psi}_k,
\end{align}
where
\begin{align}
    M_{qk}(t)
    =\begin{pmatrix}
        (\cos\alpha-1)e^{i(\tilde\epsilon_k-\tilde\epsilon_q)t} & -i\sin\alpha\, e^{-i(\tilde\epsilon_k+\tilde\epsilon_q)t}\\
        -i\sin\alpha\, e^{i(\tilde\epsilon_k+\tilde\epsilon_q)t} & (\cos\alpha-1)e^{-i(\tilde\epsilon_k-\tilde\epsilon_q)t}
    \end{pmatrix}\,,
\end{align}
with $\tilde\epsilon_k t=\epsilon_k t+\theta_k$.

Now the commutator
\begin{equation}
    \left[\Psi^\dagger_{j_0}(t)A_\alpha\Psi_{j_0}(t), \Psi^\dagger_{j}A_\alpha\Psi_{j}\right]
\end{equation}
can be calculated, see appendix \ref{app:com} for more details. We finally end up with
\begin{align}\label{anc1}
    C_{j,j_0}(t)=&\frac{1}{N^2}\sum_{k,p,k',p'}\bigg[B_{k,p}(t)B^*_{k',p'}(t)\left\langle b^\dagger_p b_kb^\dagger_{k'} b_{p'}\right\rangle\bigg]
    \nonumber\\&
    +\frac{1}{N^2}\sum_{k,p,p'}\bigg[
    A^{ba}_{k,p}(t)[A^{ba}_{k,p'}(t)]^*\left\langle b^\dagger_p b_{p'}\right\rangle
    \bigg]\,.
\end{align}
$B_{k,p}$ and $A^{ba}_{k,p}$ are as defined in appendix \ref{app:com}. We have explicitly used the fact that for the cases of interest the $a_k$ states are empty. In order to capture both the bulk and edge scrambling we must consider different Hamiltonians and dimensions. For the bulk the $a$ and $b$ states refer to the conduction and valence band, and the conduction band is assumed to be empty, \emph{i.e.}~we are in the half-filled ground state. To describe the chiral edge modes we take a one-dimensional model with counter propagating and linearly dispersing edge modes $a$ and $b$ and assume that the $a$ band is empty, leaving just a partially filled band moving in one direction. As previously mentioned this approach does leave an artifact in our results.  We can now focus on these two cases of interest.

\subsection{Bulk Scrambling}

For the bulk topological superconductor the states $b_p$ are completely filled and we find
$\langle b^\dagger_pb_kb^\dagger_{k'}b_{p'}\rangle = \delta_{kp}\delta_{k'p'}$
and $\langle b^\dagger_p b_{p'}\rangle=\delta_{pp'}$. Eq.~\eqref{anc1} then becomes
\begin{equation}\label{bulkan1}
    C_{j,j_0}(t)=\left|\frac{1}{N}\sum_kB_{k,k}(t)\right|^2
    +\frac{1}{N^2}\sum_{k,p}\left|A^{ba}_{k,p}(t)\right|^2\,.
\end{equation}
For the bulk Hamiltonian we have a slightly simpler case than the Kitaev lattice we consider numerically, as it explicitly has a chiral symmetry. We focus on the $B_{k,k}(t)$ term, which is purely from the occupied band. From appendix \ref{app:com} we find, taking the continuum limit and noting the terms in $(\cos\alpha-1)^2$ will cancel on integration over $\vec{k}$, that
\begin{equation}\label{bulkan2}
   \frac{1}{N}\sum_{\vec{k}} \frac{B_{k,k}(t)}{2\sin^2\alpha}=\textrm{Im}
   \left[ \frac{1}{N}\sum_{\vec{k}} 
    e^{i(\vec{r}-\vec{r}_0)\cdot\vec{k}+i\tilde\epsilon_kt}\right]^2\,.
\end{equation}
The positions $\vec{r}$ and $\vec{r}_0$ are those labeled by $j$ and $j_0$ respectively.

If we take, as a simple gapped model $h^x_k=M$, where $M$ is a mass inducing the gap, and $h^y_x=-2J(\cos k_x+\cos k_y)$ we can calculate $C_{j,j_0}(t)$ numerically. An example is given in appendix $\ref{app:otocs}$. Although the contributions near the top and bottom of the band result in a result which depends only on $|\vec{r}-\vec{r}_0|$ it is clear that when all contributions are taken into account the scrambling and butterfly velocity depend on the lattice and the direction of travel. This is visible in all results presented for bulk scrambling, see for example Figs.~\ref{fig:otocm0} and \ref{fig:otocsa}.

\subsection{Chiral Mode Scrambling}

For the chiral edge mode, assuming that the Fermi energy is at 0 and the Fermi momentum $p_F=0$, we have
$\langle b^\dagger_pb_kb^\dagger_{k'}b_{p'}\rangle = \theta(-p')\theta(-p)\delta_{pk}\delta_{p'k'}+\theta(-p)\theta(k)\delta_{pp'}\delta_{kk'}$ and $\langle b^\dagger_p b_{p'}\rangle=\theta(-p)\delta_{pp'}$. Eq.~\eqref{anc1} for the OTOC is now
\begin{align}
    C_{j,j_0}(t)=&\left|\frac{1}{N}\sum_{k}B_{k,k}(t)\theta(-k)\right|^2
    \nonumber\\&
    +\frac{1}{N^2}\sum_{k,p}\left|B_{k,p}(t)\right|^2\theta(-p)\theta(k)
    \nonumber\\&
    +\frac{1}{N^2}\sum_{k,p}\left|
    A^{ba}_{k,p}(t)\right|^2\theta(-p)\,.
\end{align}
We want a right moving linear dispersion and consider the Hamiltonian $\mathcal{H}=-iv_F\tau^x\partial_x$ with $\tilde\epsilon_k=-v_Fk$ and $\theta_k=0$. Details of the terms in the sums are given in appendix \ref{app:com}.

In the continuum limit performing the momentum sums as integrals with a large momentum cut-off $p_m$ results in
\begin{align}\label{resultsin}
    C_{j-j_0}(t)= &\delta_{j-j_0,v_Ft}\sin^4\alpha\nonumber\\
    &\times\left(\frac{p_m^2}{2\pi^2}-\frac{\sin^2[(j-j_0)p_m]\cos[2(j-j_0)p_m]}{\pi^2(j-j_0)^2}\right)
    \nonumber\\&
    +\left(\delta_{j-j_0,v_Ft}+\delta_{j-j_0,-v_Ft}\right)
    \sin^2\alpha(\cos\alpha-1)^2
    \nonumber\\&
    \times\left(\frac{p^2_m}{\pi^2}+\frac{p_m\sin[2(j-j_0)p_m]}{2\pi^2(j-j_0)}\right)\,.
\end{align}
The exact expression is based only on a simple model but we can extract the essential behavior. Neglecting the artifact of a left moving term generated by perturbing the empty left moving states we have
\begin{equation}\label{chiral_otoc}
    C_{j-j_0}(t)\sim\delta_{j-j_0,v_Ft}\left(a
    +\frac{c}{(j-j_0)^2}\right)\,,
\end{equation}
with $a$ and $c$ some terms depending on the details of the model and perturbation. The OTOC contains a constant term which moves along the edge at the velocity $v_F$, without being scrambled - a dynamical scar. This behavior is confirmed by the numerical results of the following section, see Figs.~\ref{fig:otocwf} and \ref{fig:edge_speed}. The validity of this calculation relies on there being a large enough topological gap such that we can treat a linear dispersing mode independently. Reducing the topological gap will reduce the weight of the unscrambled OTOC, as can be seen from Eq.~\eqref{resultsin}.

\begin{figure*}
    \centering
    \includegraphics[width=0.9\textwidth]{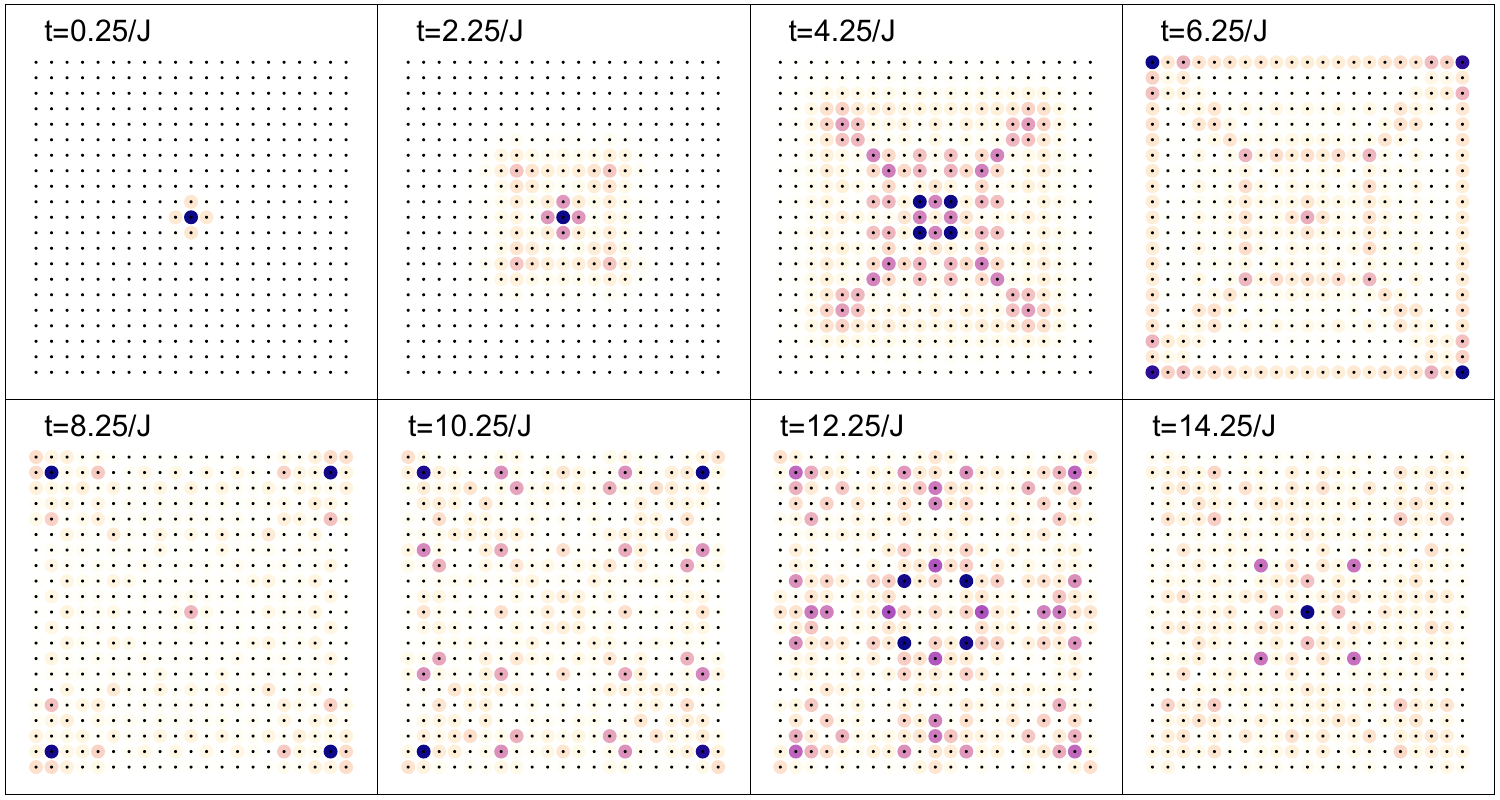}\\
    \includegraphics[width=0.35\textwidth]{OTOC_Legend.pdf}
    \caption{The OTOC $C_{j,j_0}(t)$ at different time steps following a perturbation in the center, at site $\vec{r}_0=(11,11)a$, in the topologically non-trivial phase with $\nu=1$, see main text and fig.~\ref{fig:otocm0} for more details. After a time $t\approx 6/J$ the correlations have hit the edge of the system and scatter back, resulting in a quickly scrambled system.}
    \label{fig:otocm1}
\end{figure*}

\section{Scrambling in the Bulk}\label{bulk}

In this section we present the main results of the numerical calculations of the OTOCs, comparing them where possible to analytical results, for the Kitaev lattice. As a first check we consider the scrambling behavior in the bulk of a topologically trivial system, see Fig.~\ref{fig:otocm0}. We plot results for system sizes $21\times21$. As expected the initially localized OTOC spreads throughout the system. We note that, as predicted by Eqs.~\eqref{bulkan1} and \eqref{bulkan2}, the velocity with which the scrambling occurs depends on the direction. Let us note here some convenient nomenclature, we define $C_{\rm max}(t)=\textrm{max}[C_{j,j_0}(t)]$ and $C_{\rm max}=\textrm{max}[C_{\rm max}(t)]$. \emph{I.e.}~we define a maximum at a particular time for all positions, and a global maximum of the OTOC for all times. We use these for normalizing the data optimizing for presentation.

\begin{figure}
    \centering
    \includegraphics[width=0.9\columnwidth]{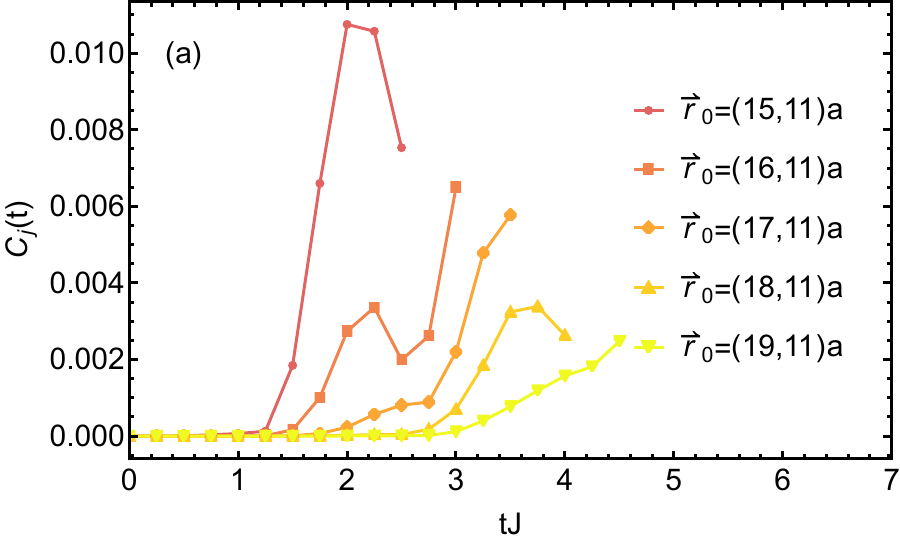}
    \includegraphics[width=0.9\columnwidth]{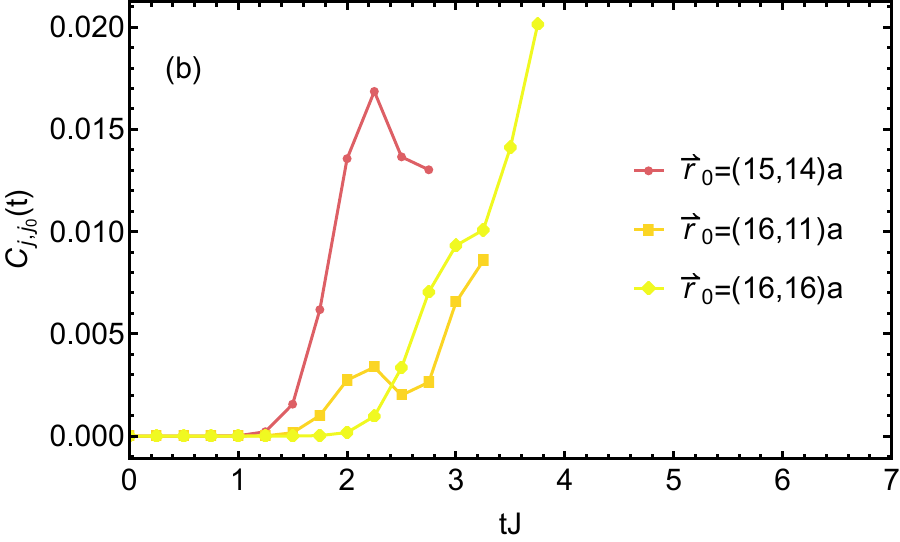}
    \caption{The OTOC $C_{j,j_0}(t)$ as a function of $t$ for different sites $\vec{r}$ following a perturbation in the center at site $\vec{r}_0=(11,11)a$. This is for the topologically non-trivial phase with $\nu=1$.  In the panel (a) one can see that as we move further from the perturbation along the $x$-direction the onset of the non-zero OTOC moves later in time, as expected. In panel (b) we compare several different points, two are an equal distance from the perturbation along different directions, one is further away along the diagonal.}
    \label{fig:otoct1}
\end{figure}

In Fig.~\ref{fig:otocm1} we plot results for a perturbation in the center of the system in a topologically non-trivial phase. In the bulk scrambling the topological phase plays no significant role and we see similar results to Fig.~\ref{fig:otocm0}. One can note that details of how fast the scrambling spreads and in which direction depend on the details of the band structure. In principle we may expect an effect from the chiral edge modes once the OTOC is non-zero at the boundary, but any effect is too small to see in this case. We focus on these effects in Sec.~\ref{chiral}.

To demonstrate the directional dependence of the scrambling, and the existence of a butterfly velocity more clearly in Fig.~\ref{fig:otoct1} we plot $C_{j,j_0}(t)$ as a function of $t$ for different points at different distances and along different directions from the site of the perturbation $\vec{r}_0$. Fig.~\ref{fig:otoct1}(a) demonstrates the existence of a butterfly velocity, although complications of the two-dimensional case prevent us from satisfactorily fitting an effective formula through these points. What is clearly visible is that the OTOC becomes appreciably non-zero at a time which increases monotonically with the distance form the perturbation.

Fig.~\ref{fig:otoct1}(b) considers three different points near the perturbation. Two are at an equal distance: $\vec{r}-\vec{r}_0=(4,3)a$ and $\vec{r}-\vec{r}_0=(5,0)a$. The third is $\vec{r}-\vec{r}_0=(5,5)a$ which is located along a diagonal. We see by comparison of the first two that there is some dependence on direction for the butterfly velocity, which is further confirmed by comparison with the third point.

\section{Scrambling with Chiral Edge Modes}\label{chiral}

Our main motivation for studying OTOCs in two-dimensional topological models is to see how the topologically protected edge modes affect the scrambling. We now focus on this question, placing the perturbation on the edge of our lattice. In this section we again focus on the Kitaev lattice with chiral edge modes. In Sec.~\ref{helical} we will address what happens when helical edge modes are present.

\begin{figure*}
    \centering
    \includegraphics[width=0.9\textwidth]{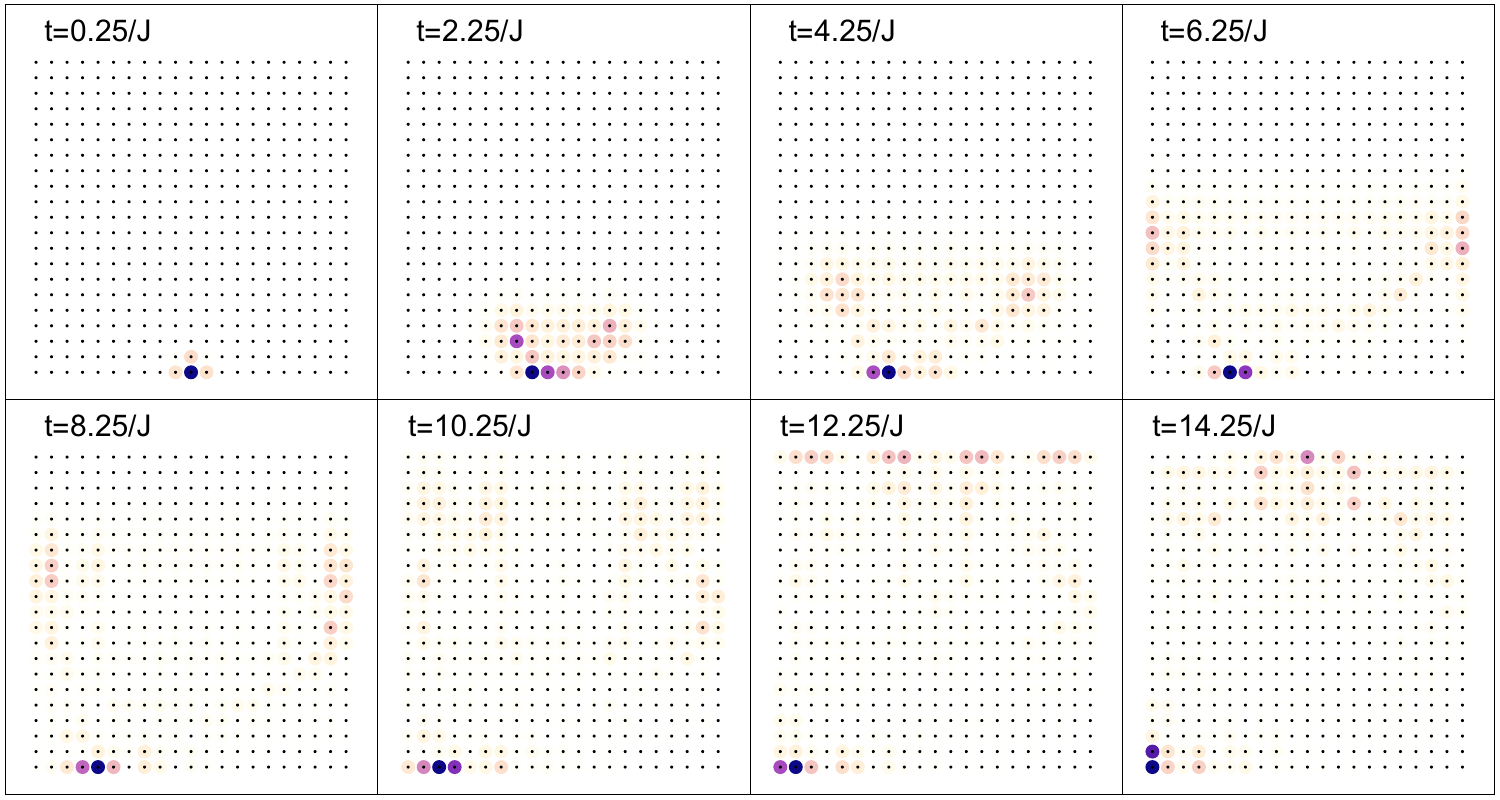}\\
    \includegraphics[width=0.35\textwidth]{OTOC_Legend.pdf}
    \caption{The OTOC $C_{j,j_0}(t)$ at different time steps following a perturbation on the edge, at site $\vec{r}_0=(1,11)a$, in the topologically non-trivial phase with $\nu=1$, see main text and fig.~\ref{fig:otocm0} for more details. In the $\nu=1$ phase the chiral edge modes propagate clockwise, as does the scar in the scrambling visible as a dark purple region on the boundary.}
    \label{fig:otoc1}
\end{figure*}

\begin{figure*}
    \centering
    \includegraphics[width=0.9\textwidth]{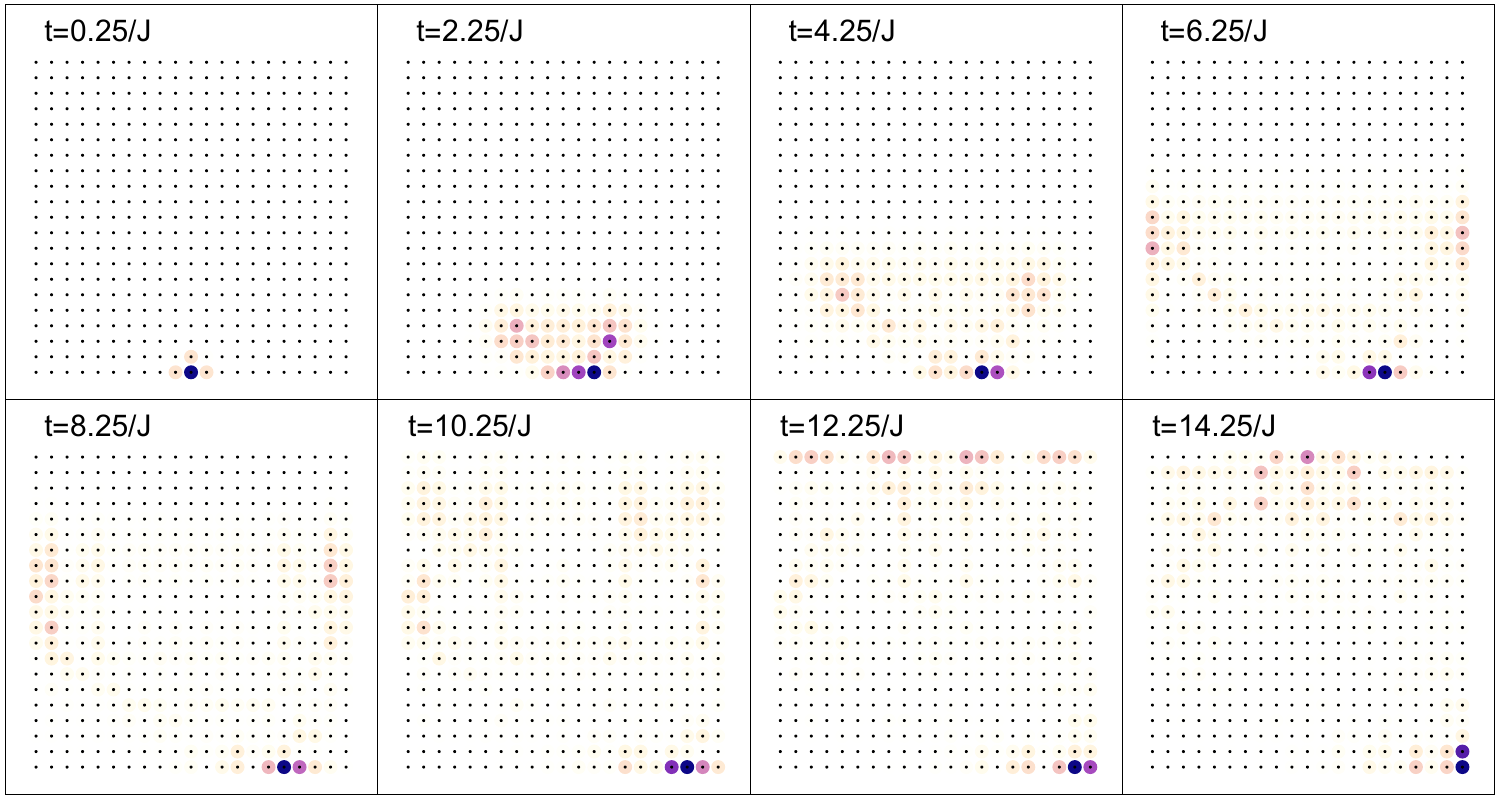}\\
    \includegraphics[width=0.35\textwidth]{OTOC_Legend.pdf}
    \caption{The OTOC $C_{j,j_0}(t)$ at different time steps following a perturbation on the edge, at site $\vec{r}_0=(1,11)a$, in the topologically non-trivial phase with $\nu=-1$, see main text and fig.~\ref{fig:otocm0} for more details. In the $\nu=-1$ phase the chiral edge modes propagate counter-clockwise, as does the scar in the scrambling visible as a dark purple region on the boundary.}
    \label{fig:otocminus1}
\end{figure*}

First we note that if we perturb the topologically trivial system at the edge we notice no particular effect at the boundary, see Fig.~\ref{fig:otoc0} in appendix \ref{app:otocs}. One sees the scrambling spreading throughout the system with an additional effect caused by scattering from the nearby boundary. This is in stark contrast to the topologically non-trivial phases, see Figs.~\ref{fig:otoc1} and \ref{fig:otocminus1}.

In Fig.~\ref{fig:otoc1} we consider the phase $\nu=1$, where the chiral modes propagate clockwise. The existence of an unscrambled contribution to the OTOC, moving around the boundary clockwise, exactly as the chiral modes propagate is evident. This ``scar'' moves around the corner without backscattering, as would be expected if it is carried by the topologically protected edge modes. If we reverse the propagation of the edge modes to counter-clockwise by changing the sign of the Chern number to $\nu=-1$, then we find the scar propagates counter-clockwise. See Fig.~\ref{fig:otocminus1} for confirmation. We have also checked that if we stay in the $\nu=1$ phase but change the sign of $\Delta$ we still see clockwise propagation for both the edge modes and the scar in the scrambling as expected.

\begin{figure}
    \centering
    \includegraphics[width=0.9\columnwidth]{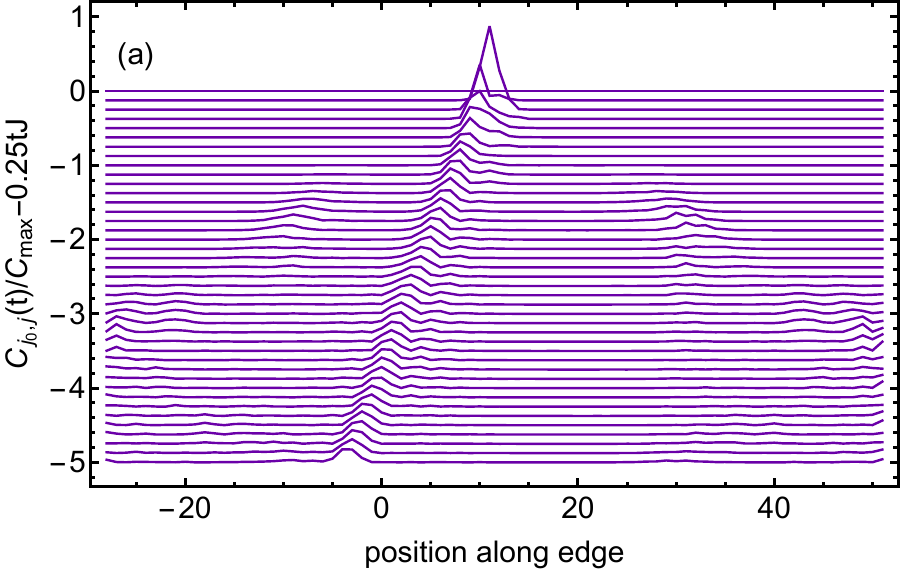}\\
    \includegraphics[width=0.9\columnwidth]{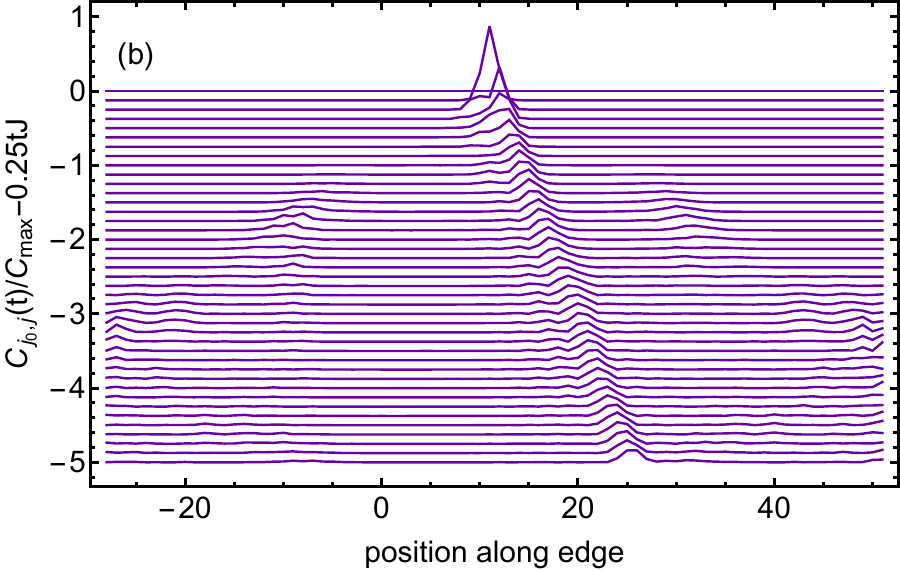}
    \caption{Waterfall plots for the OTOC only on the edge of the lattice, plotted along the $x$-axis. Each time step shown is offset as given on the $y$-axis. Results are for $\nu=1,-1$ from top to bottom. In the two topologically non-trivial phases a clearly visible scar is moving around the edge with, after a transient decay, a constant magnitude as predicted by Eq. \eqref{chiral_otoc}.}
    \label{fig:otocwf}
\end{figure}

This dynamical scar is propagated for long times by the edge states in their chiral direction without noticeable decay. This can not be seen directly in Figs.~\ref{fig:otoc1} and \ref{fig:otocminus1} due to the necessary time dependent normalization of the color scheme.  Fig.~\ref{fig:otocwf} plots the OTOC only on the boundary sites. Here we have normalized the OTOC by a single scaling factor for all times and plot the results as a waterfall plot. Both the constant velocity and the constant magnitude, after a short transient time, of the OTOC scar are evident. We do not see any decay of the scar over the accessible timescales. We note that for these results, in contrast to the analytical calculation, non-linear effects of the dispersion relation of the edge modes and the bulk modes are all present. For the topologically trivial phase only small effects from the bulk scrambling hitting the boundary are visible, see Fig.~\ref{fig:otocwfapp} in appendix \ref{app:otocs}.

\begin{figure}
    \centering
    \includegraphics[width=0.9\columnwidth]{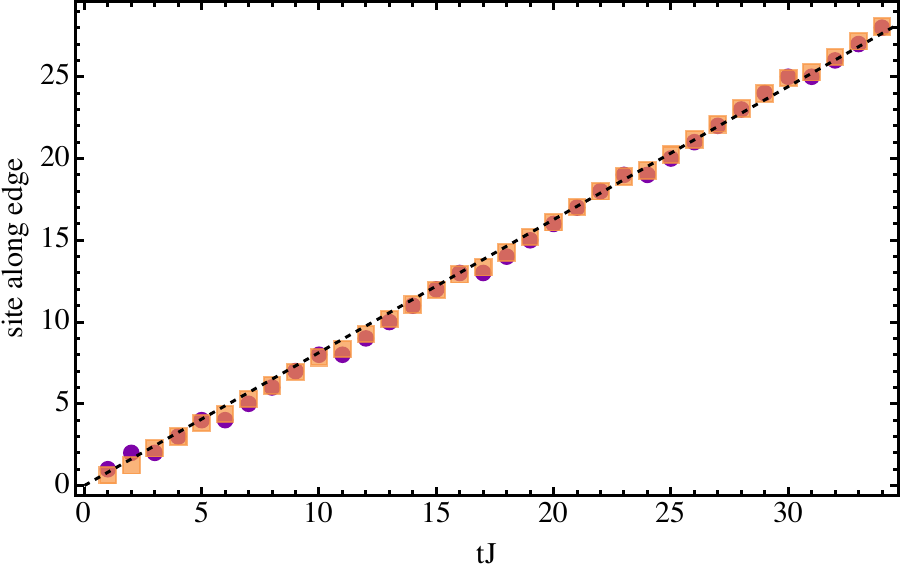}
    \caption{The location of the dynamical scar on the boundary, enumerated by site along the edge with $j=0$ the location of the perturbation. For the topologically non-trivial phase with $\nu=1$. Circles show the average of the peak and its two nearest neighbors, and the squares show the center of mass. Both agree well with each other for the position of the scar. The velocity is extracted via a linear fit and gives $v=0.813aj$. The velocity of the linear edge modes is $v=0.896aj$.}
    \label{fig:edge_speed}
\end{figure}

To test the prediction that for a linearly dispersing chiral mode the scar should move with the velocity of the modes themselves we extract the position of the OTOC peak on the boundary as a function of time. To do this we consider two possible methods for determining the location of the scar. First we take the average position of the peak and its 2 closest neighbors on each side, second we take a weighted average of the whole non-zero region of the peak - \emph{i.e.}~the center of mass of the scar. We find reasonable agreement between the velocities with the velocity extracted from the peak being $v=0.813aj$ and the velocity of the linear edge modes $v=0.896aj$, see Fig.~\ref{fig:edge_speed}. In the real lattice model the linear modes are not perfectly linear except at very low energies, which causes the finite width of the peak compared to the theoretical prediction fo a perfectly sharp peak.

\begin{figure*}
    \centering
    \includegraphics[width=0.95\textwidth]{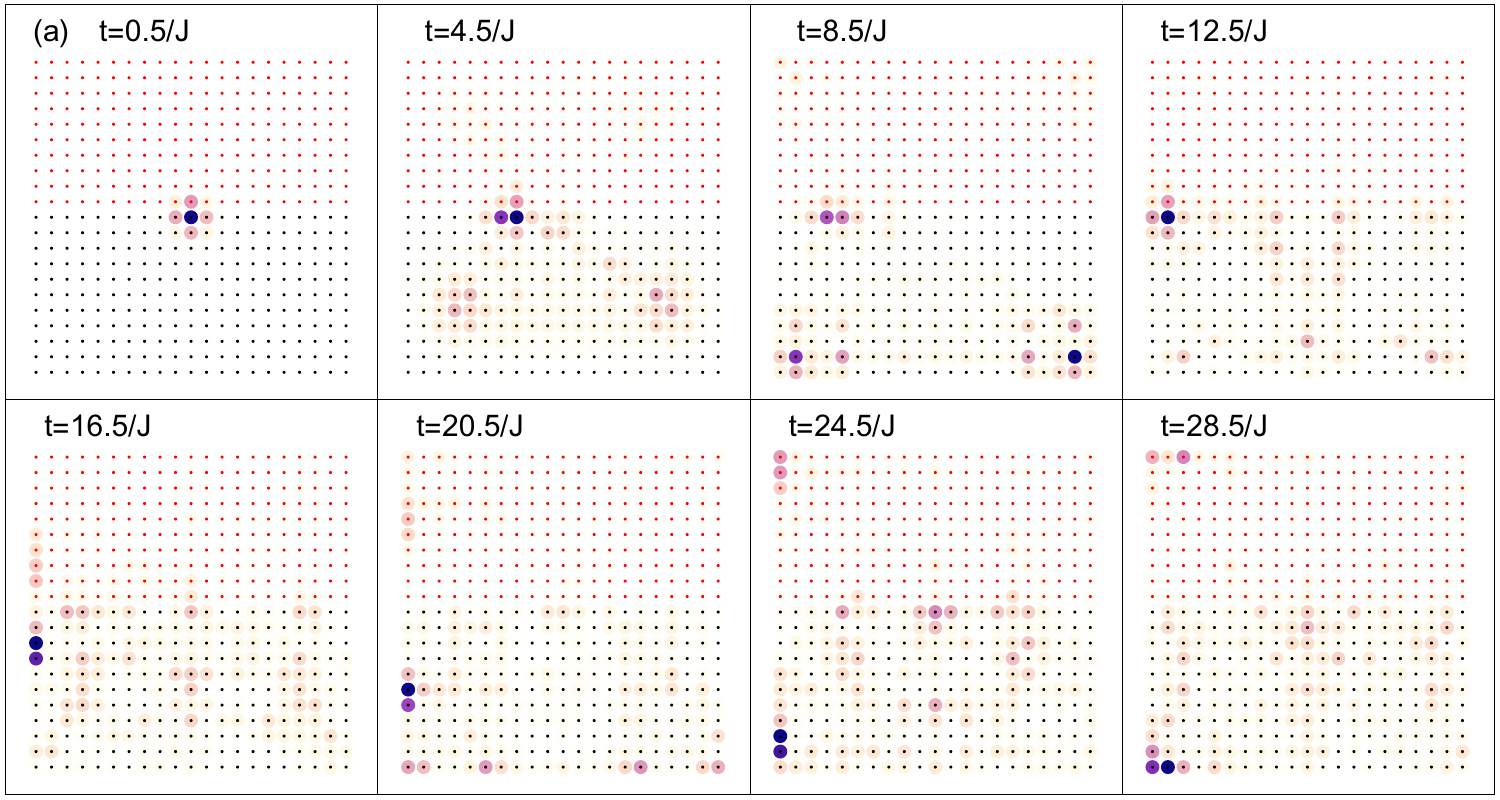}\\
    \includegraphics[width=0.95\textwidth]{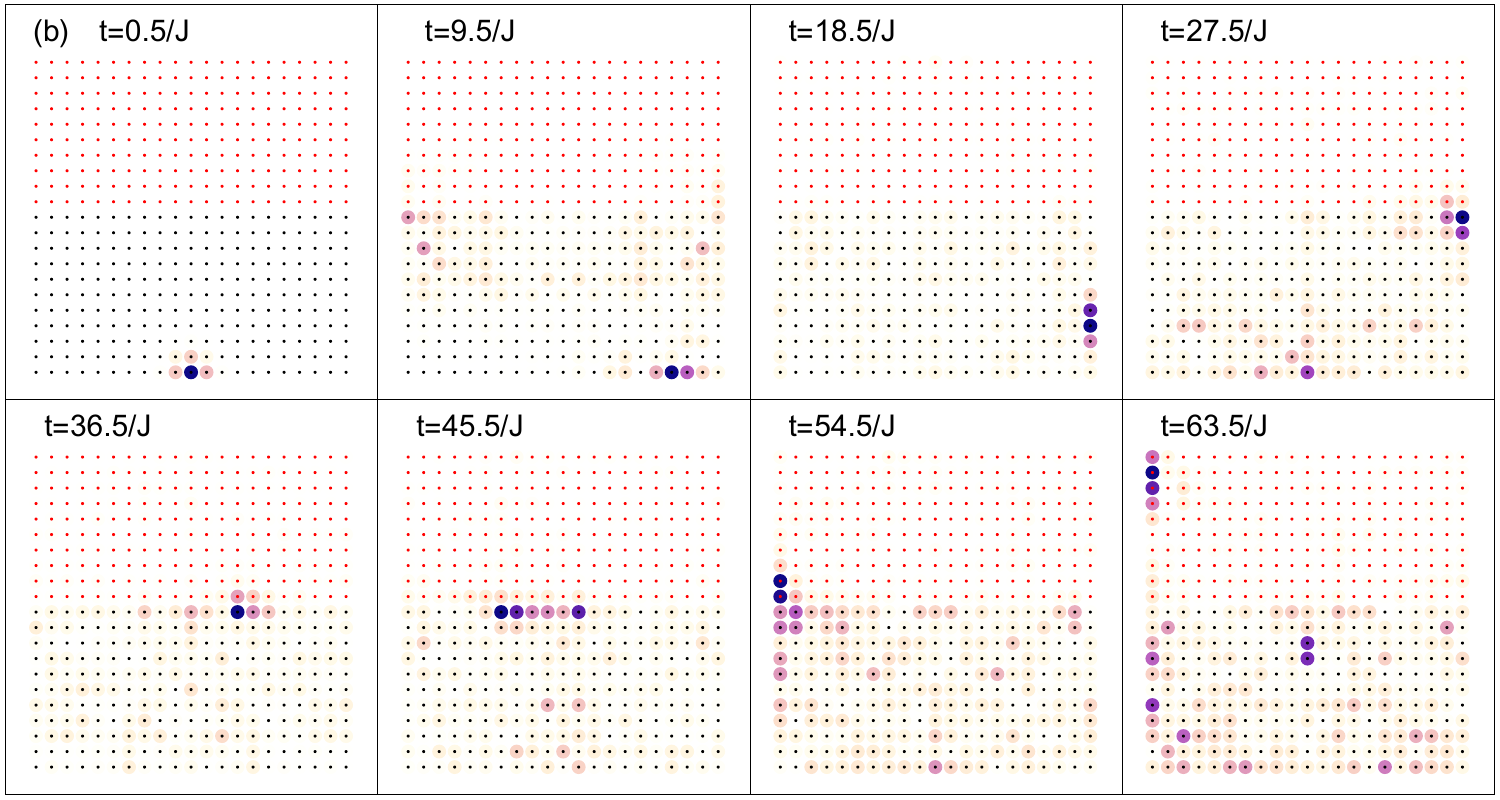}\\
    \includegraphics[width=0.35\textwidth]{OTOC_Legend.pdf}
    \caption{The OTOC $C_{j,j_0}(t)$ at different time steps in a system where there is a domain wall between $\nu=-1$ and $\nu=1$ such that for $y>11a$ $\nu=1$ and for $y<12a$ $\nu=-1$. Each phase is marked by a different color for the lattice site. See the main text and fig.~\ref{fig:otocm0} for more details. In panel (a) the perturbation is at $\vec{r}_0=(11,11)a$ located just inside the lower domain. The scar is moving as predicted, with a small weight moving in the upper domain in the opposing direction as necessary. If we place the perturbation at $\vec{r}_0=(1,11)a$, as in panel (b), then it becomes mixed along the domain wall into two contributions traveling clockwise and counter-clockwise in the two domains.}
    \label{fig:otocdw}
\end{figure*}

As a final example we take a lattice with a domain wall between two different topological phases with $\nu=1$ and $\nu=-1$. The lattice size is $21\times21$ as always and we place the domain wall between sites 11 and 12 along the $y$-direction so that for $y>11a$ $\nu=1$ and for $y<12a$ $\nu=-1$, see Fig.~\ref{fig:otocdw}. We then place $\vec{r}_0$ at two different locations. As may be expected if the scar is caused by a lack of scrambling in the linearly dispersing edge modes we see the scar moving around the topological boundary. Along the domain wall the scar can be mixed between the modes of the two phases and moves on along both boundaries when they become separated. This occurs whether the original perturbation was at the domain wall or along the boundary of a region of a single topological phase.

\section{Scrambling with Helical Edge Modes}\label{helical}

\begin{figure*}
    \centering
    \includegraphics[width=0.95\textwidth]{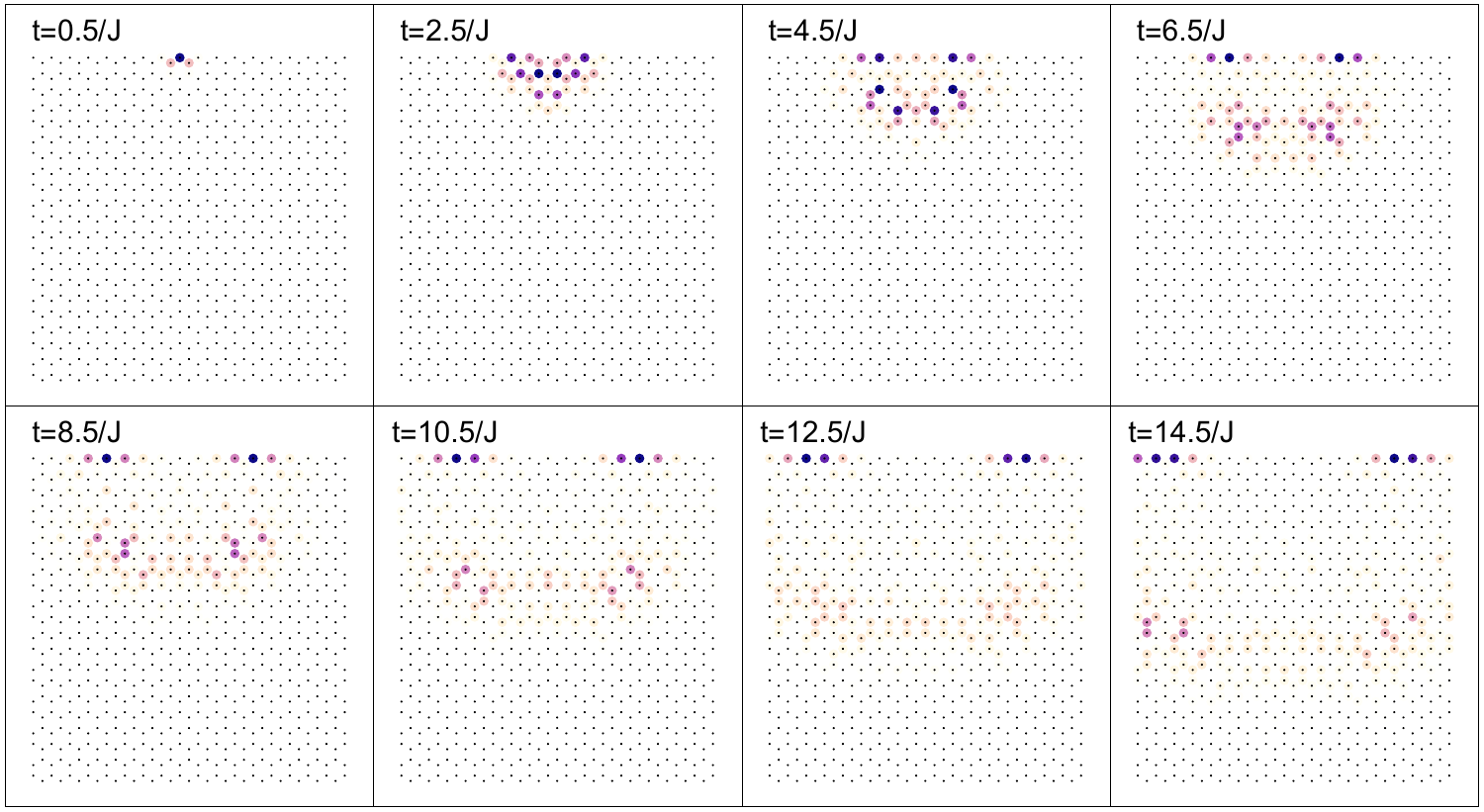}\\
    \includegraphics[width=0.35\textwidth]{OTOC_Legend.pdf}\\
    \caption{The OTOC $C_{j,j_0}(t)$ at different time steps following a perturbation on the top edge in the topologically non-trivial phase of a $\mathbb{Z}_2$ topological insulator. In this phase counter propagating helical edge modes are present, see main text and fig.~\ref{fig:otocm0} for more details. As we predict two counter-propagating scars can be seen.}
    \label{fig:otocz2}
\end{figure*}

Topological insulators and superconductors can also have multiple bands inside the gap, with different directions. Here we consider the main additional feature this can bring: counter-propagating modes. In this section we therefore turn to the $\mathbb{Z}_2$ topological insulator \eqref{km_ham}. The expectation based on the previous results is that we have counter propagating dynamical scars, and we see this borne out in Fig.~\ref{fig:otocz2}.

\begin{figure*}
    \centering
    \includegraphics[width=0.95\textwidth]{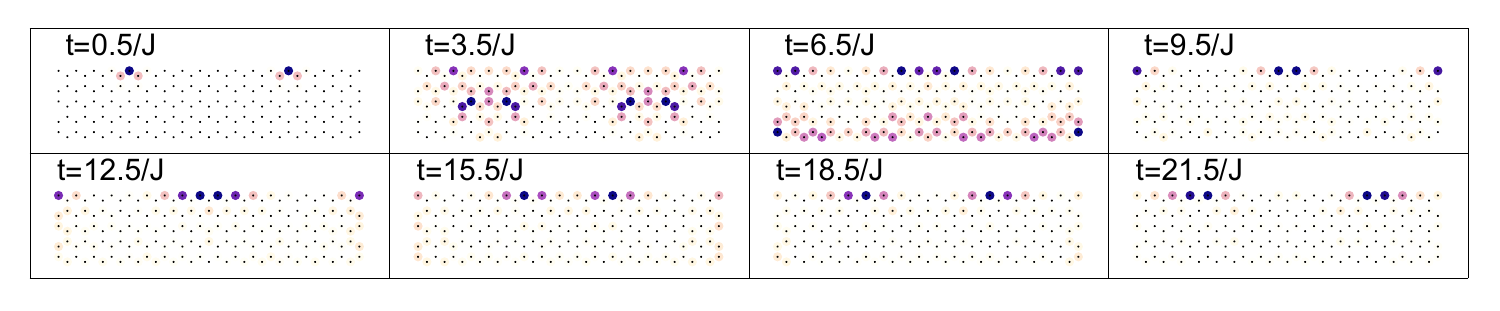}\\
    \includegraphics[width=0.35\textwidth]{OTOC_Legend.pdf}\\
    \caption{The OTOC $C_{j,j_0}(t)$ at different time steps following a perturbation located at two sites on the top edge, see Eq.~\eqref{twopert}. This is for the topologically non-trivial phase of a topological insulator with counter propagating helical edge modes, see main text and fig.~\ref{fig:otocm0} for more details. We see the counter propagating scars which pass through each other and continue moving along the edge unaffected. Here we show a portion of the larger lattice with just the top edge visible.}
    \label{fig:otocz22}
\end{figure*}

As the Kane-Mele model \eqref{km_ham} is based on a honeycomb lattice Fig.~\ref{fig:otocz2} also demonstrates that we do not need a smooth edge for the scar to propagate as it is carried by the topologically protected edge modes. Further examples of this can be seen in Fig.~\ref{fig:otocapp} in appendix \ref{app:otocs}. One natural question which does arise if we have multiple dynamical scars is whether they interact with each other, leading to scrambling. In Fig.~\ref{fig:otocz22} we plot the results for the OTOC following a perturbation at two sites:
\begin{equation}\label{twopert}
    \hat V_{j_0,j_1}=
    e^{i\alpha\Phi_{j_0}^\dagger\tau^z\Phi_{j_0}+i\alpha\Phi_{j_1}^\dagger\tau^z\Phi_{j_1}},
\end{equation}
$\hat W_j$ remains unchanged. For convenience we will retain the notation $C_{j,j_0}(t)$ for the OTOC, although it now also depends on $j_1$ which we do not vary. Fig.~\ref{fig:otocz22} demonstrates that the two sites of the perturbations each generate counter propagating scars. When the two scars approaching each other ``meet'' they pass straight through each other, consistent with their origin in the topologically protected helical edge modes. Thus this generates no scrambling of these scars.

\section{Discussion and Conclusions}\label{conclusion}

In this article we have studied scrambling in two dimensional topological insulators and superconductors. Both analytical and numerical results have been based on simple exemplary models, and one natural question to ask is to what extent the results we have seen generalize to more complicated many-band topological models. We have studied scrambling through the lens of the OTOC, which characterizes the way in which information is spread throughout the system during unitary time evolution, becoming unavailable for any practical resurrection. For the bulk of the topological models we see no significant difference to other gapped models, and no direct effect of the band topology. One distinction is that the butterfly velocity becomes direction dependent as the entire band structure is important for scrambling, rather than just a higher symmetry low energy region of it.

Most interesting is the effect on the edges when there are topologically protected edge modes present. We see dynamical scarring moving with the speed and direction of the topologically protected chiral or helical, as appropriate, edge modes. The dynamical scar does not undergo scrambling over the time scales we investigated, and we derived an analytical formula based on a simple model of a linearly dispersing edge mode. The scars are impervious to each other, and do not scramble when meeting, but pass through each other unnoticed. We see also no effect from more complicated shapes of the edges. We do note that domain walls which contain multiple edge modes allow for possible mixing of the scars between different modes propagating in the same direction, but they seem to still remain unscrambled.

These results suggest some possible further studies. The effects of disorder, in particular symmetry breaking disorder which may introduce scrambling into the scars could be investigated. Additionally to what extent these results will remain true in systems with many different edge modes present, due for example to higher Chern numbers, could be of interest. Can co-propagating modes introduce dissipation where counter-propagating ones could not?

The role of interactions and relaxation is another open question. We expect that the dynamical scarring should survive for weak interactions which do not close the topological gap. However to determine whether this is the case would require a detailed study which we leave for future work. One may also check whether temperature has a significant role to play.

We also note that the scars allow for a dynamical probe of the edge modes, and could be of interest in dynamical situations when simpler correlation functions are not suitable probes of the presence of topologically protected edge modes.

\acknowledgments
N.S.~ and S.G.~acknowledge the support from the National Science Centre (NCN, Poland) under the grant 2024/53/B/ST3/02600. All data used in this article, some supporting data, and also videos of the scrambling, can be found at Ref.~\cite{Szpara2025}. The code for calculating the scrambling in the Kitaev lattice is available at \href{https://github.com/Dominik-S-code/Topological-Scrambling-2D}{https://github.com/Dominik-S-code/Topological-Scrambling-2D}.

\appendix

\counterwithin{figure}{section}

\section{Further examples of the OTOCs}\label{app:otocs}

\begin{figure*}
    \centering
    \includegraphics[width=0.9\textwidth]{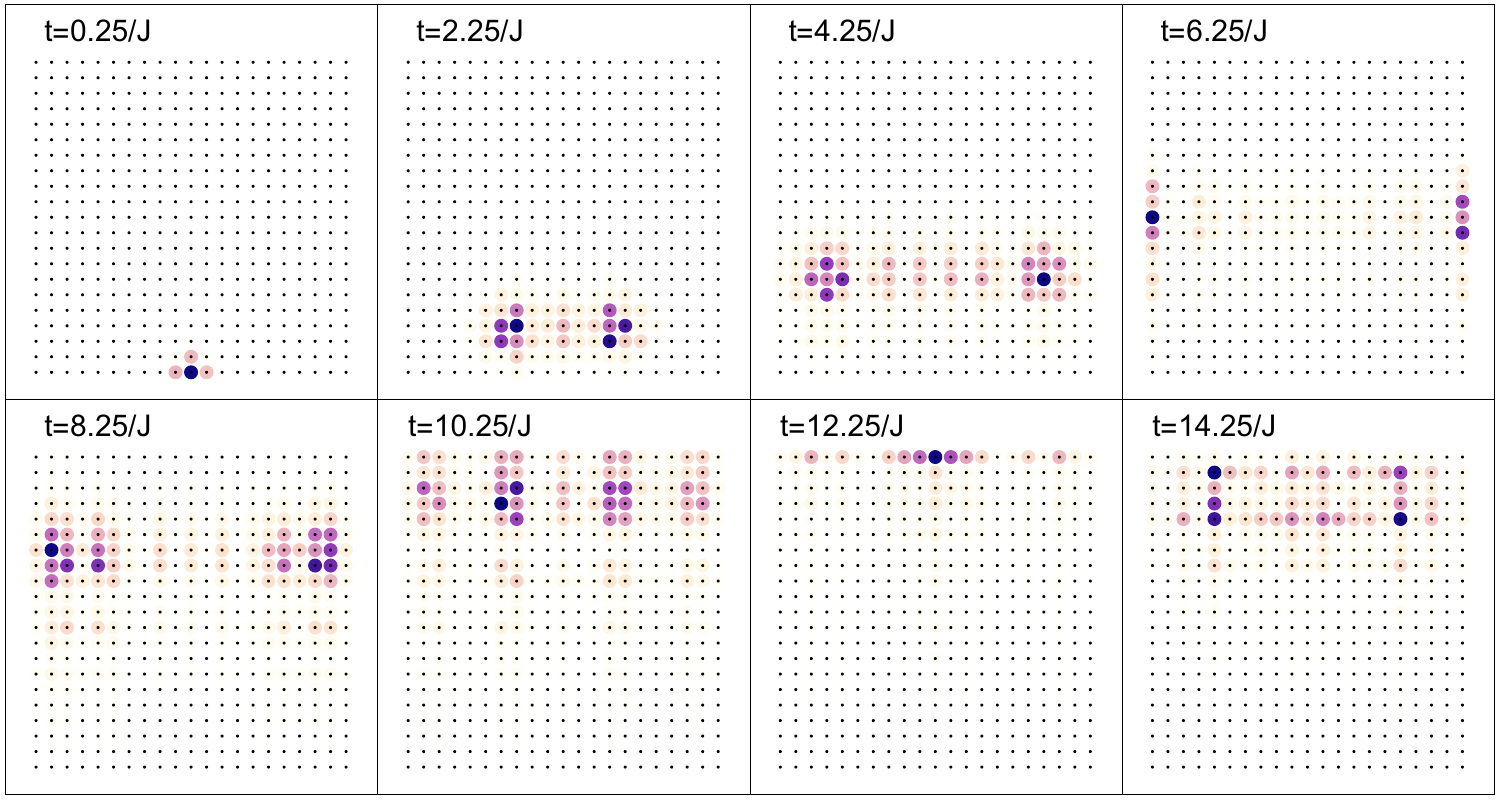}\\
    \includegraphics[width=0.35\textwidth]{OTOC_Legend.pdf}
    \caption{The OTOC $C_{j,j_0}(t)$ at different time steps following a perturbation on the edge, at site $\vec{r}_0=(1,11)a$, in the topologically trivial phase with $\nu=0$, see main text and fig.~\ref{fig:otocm0} for more details. No chiral edge modes are present and we see no scar in the scrambling visible on the boundary.}
    \label{fig:otoc0}
\end{figure*}

In this appendix we present some more examples of the numerical and analytical OTOC calculations. In Fig.~\ref{fig:otoc0} we show the results for an initial perturbation on the edge of a topologically trivial system. As expected there is no scarring generated on the edge, with the initial perturbation scrambling into the bulk.

\begin{figure}
    \centering
    \includegraphics[width=0.9\columnwidth]{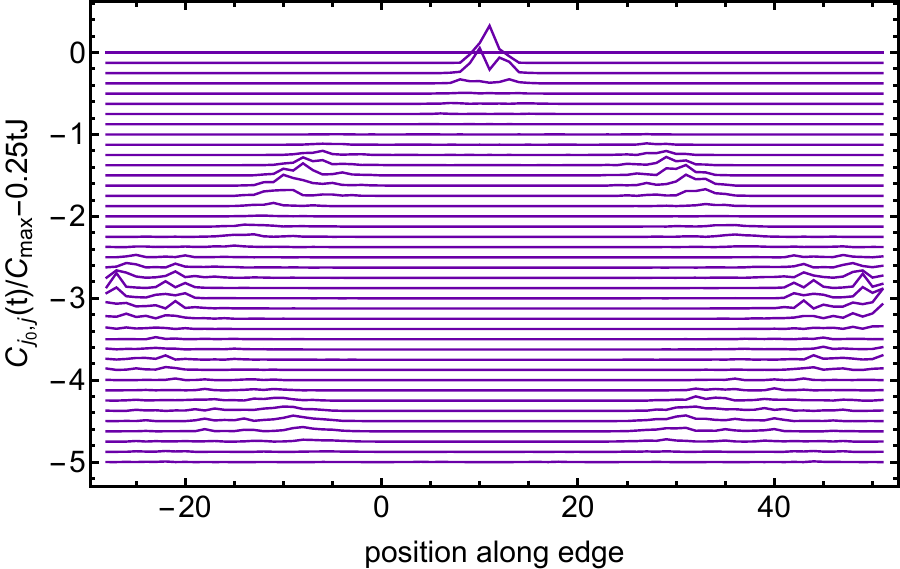}
    \caption{Waterfall plot for the OTOC only on the edge of the lattice, plotted along the $x$-axis. Each time step shown is offset as given on the $y$-axis. Results are for $\nu=0$. The perturbation moves away from the boundary, with later time effects caused by the bulk scattering off the boundary.}
    \label{fig:otocwfapp}
\end{figure}

In Fig.~\ref{fig:otocwfapp} we plot the edge of the topologically trivial lattice following a perturbation on the edge as a waterfall plot. This confirms that no scarring occurs on the edge in this scenario.

\begin{figure*}
    \centering
    \includegraphics[width=0.9\textwidth]{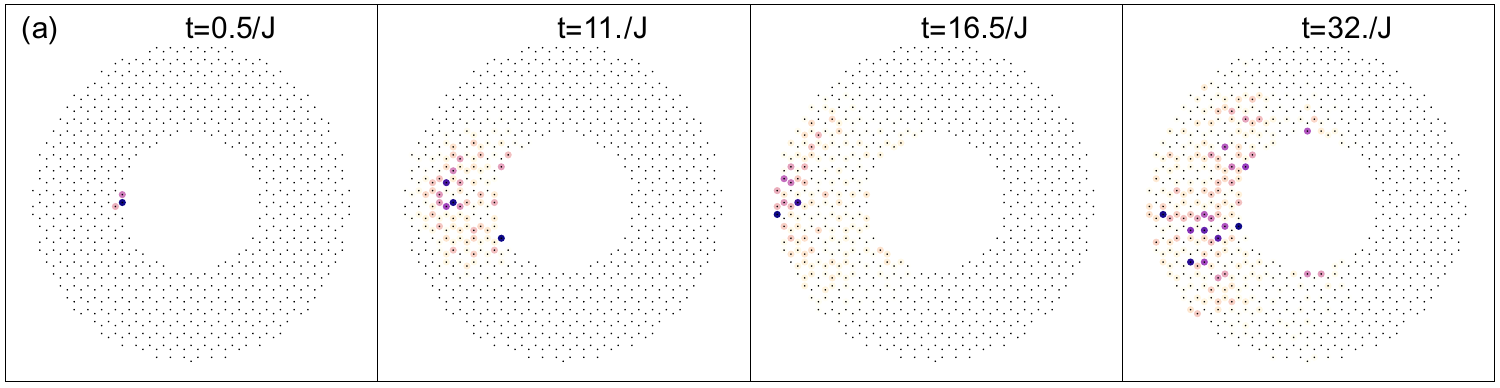}\\
    \includegraphics[width=0.9\textwidth]{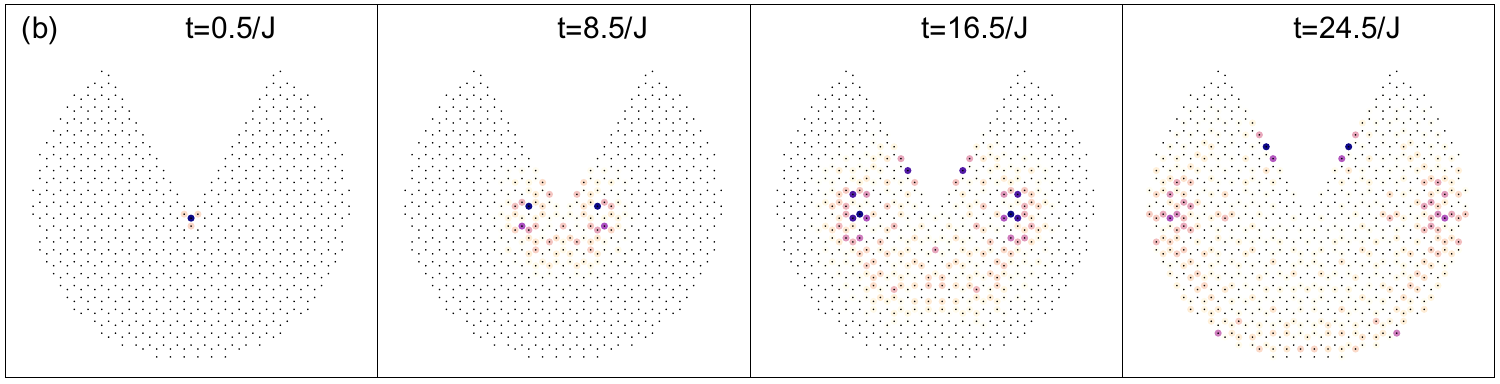}\\
    \includegraphics[width=0.35\textwidth]{OTOC_Legend.pdf}
    \caption{The OTOC $C_{j,j_0}(t)$ at different time steps for the topologically non-trivial phase of a $\mathbb{Z}_2$ topological insulator with counter propagating helical edge modes, see main text and fig.~\ref{fig:otocm0} for more details. Locations of the perturbations can be seen in the early time plots. (a,b) show two different shapes for the same parameters demonstrating the scars propagate regardless of details of the geometry of the boundary. In panel (a) at intermediate times the OTOC spreading to the outer edge has a larger weight and is therefore more easily visible than the OTOC scar moving around the inner edge.}
    \label{fig:otocapp}
\end{figure*}

In Fig.~\ref{fig:otocapp} we show several more examples of the scrambling for $\mathbb{Z}_2$ topological insulator with helical edge modes. Here we show systems with more complicated boundaries, demonstrating that the scarring is still propagating around the edges.

\begin{figure*}
    \centering
    \includegraphics[width=0.9\textwidth]{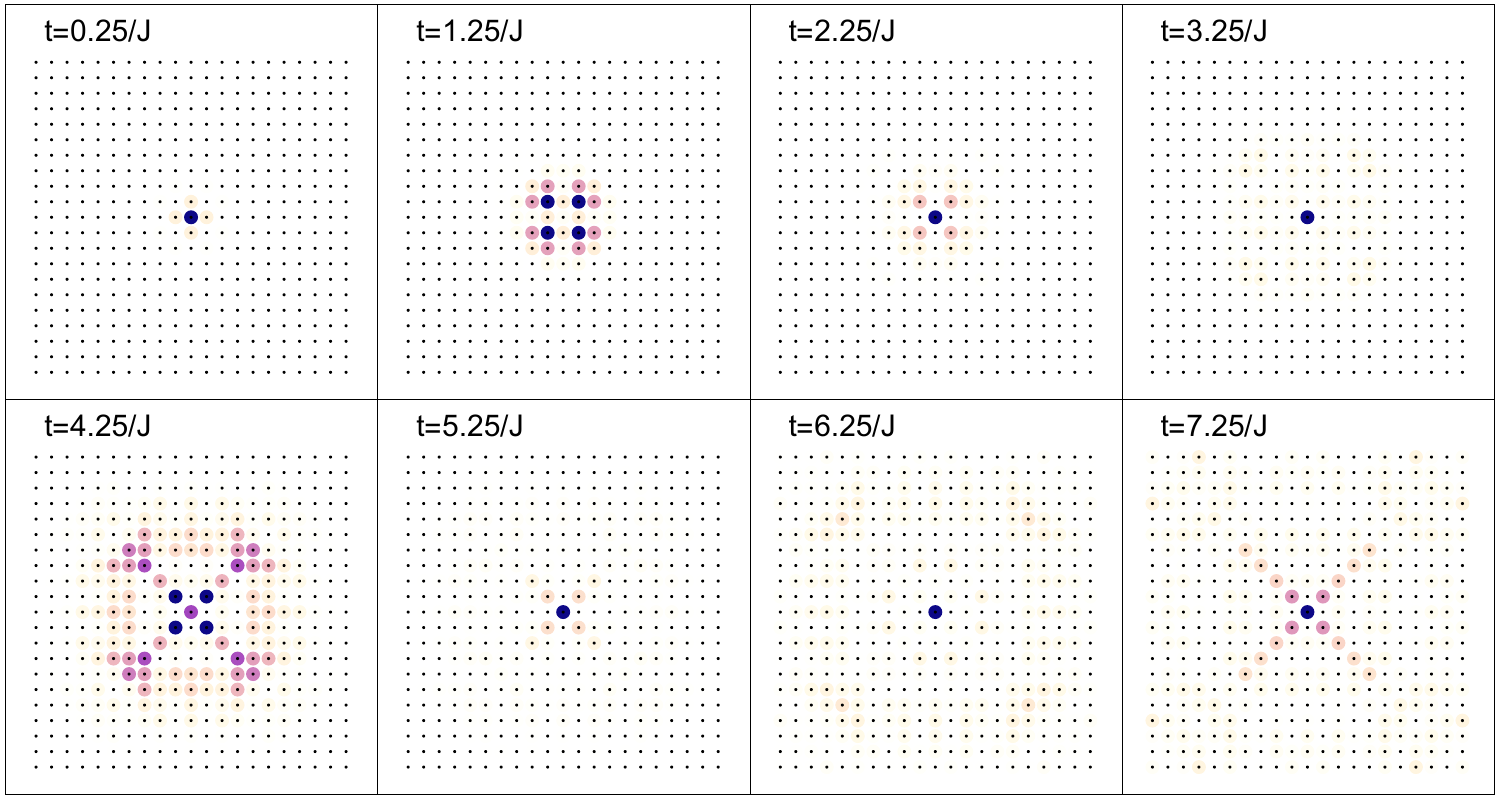}\\
    \includegraphics[width=0.35\textwidth]{OTOC_Legend.pdf}
    \caption{The semi-analytical OTOC $C_{j,j_0}(t)$ from Eqs.~\eqref{bulkan1} and \eqref{bulkan2} at different time steps following a perturbation in the center, at site $\vec{r}_0=(11,11)a$, in the topologically trivial phase with $\nu=0$. See main text for details and fig.~\ref{fig:otocm0} for more details. Here we check the bulk scrambling which spreads throughout the system with a characteristic velocity, dependent on direction, see Sec.~\ref{sec_analytics} for details. We can note a term at the central site which oscillates and decays in magnitude, and spreading correlations along different directions. We note again that the normalization for the color scheme is performed at each time step separately, or little would be visible.}
    \label{fig:otocsa}
\end{figure*}

In Fig.~\ref{fig:otocsa} we plot results of numerically integrating Eqs.~\eqref{bulkan1} and \eqref{bulkan2}. The mass is taken to be $M=2J$ and the perturbation is at the center of the lattice, $\vec{r}_0=(11,11)a$. As we see in other examples the initial perturbation is scrambled, spreading out into the bulk with a different velocity along different directions. Additionally one can see the oscillations in the magnitude of the scrambling which occur. We have further considered a variety of parameters at different time steps to check the generality of these conclusions.

\section{Commutators}\label{app:com}

In this appendix we give some further details of the analytical calculations. For the calculation of the OTOC we require the commutator
\begin{equation}
    \left[\Psi^\dagger_{j_0}(t)A_\alpha\Psi_{j_0}(t),\Psi^\dagger_{j}A_\alpha\Psi_{j}\right]
\end{equation}
has multiple terms. However we will consider only those which will contribute when averaged over a state which contains no $a$ states. The justification being that $b$ corresponds in the bulk to the filled negative energy band, and for the linear edge modes to one direction for the chiral edge states. The only terms left which will contribute are therefore
\begin{equation}
    \frac{1}{N}\sum_{k,p}
    \left(B_{k,p}(t)b^\dagger_p b_k+A^{ba}_{k,p}(t)b^\dagger_pa_k+A^{ab}_{k,p}(t)a^\dagger_pb_k\right)\,.
\end{equation}
The coefficients of these terms are:
\begin{align}
    B_{k,p}(t)=&(\cos\alpha-1)^2\nonumber\\&\times\frac{1}{N}\sum_{q}
    \bigg(e^{ij_0(q-p)+ij(k-q)+i(\tilde\epsilon_p-\tilde\epsilon_q)t}\nonumber\\&
    -e^{ij_0(k-q)+ij(q-p)+i(\tilde\epsilon_q-\tilde\epsilon_k)t}\bigg)
    \nonumber\\+
    &\sin^2\alpha\frac{1}{N}\sum_{q}
    \bigg(e^{ij_0(k-q)+ij(q-p)+i(\tilde\epsilon_k+\tilde\epsilon_q)t}\nonumber\\&
    -e^{ij_0(q-p)+ij(k-q)-i(\tilde\epsilon_q+\tilde\epsilon_p)t}\bigg)\,,
\end{align}
along with
\begin{align}
    A^{ba}_{k,p}(t)=&i\sin\alpha(\cos\alpha-1)
    \nonumber\\&\times\frac{1}{N}\sum_{q}\bigg(
    e^{ij_0(k-q)+ij(q-p)-i(\tilde\epsilon_q-\tilde\epsilon_k)t}
    \nonumber\\&
    +e^{ij_0(k-q)+ij(q-p)-i(\tilde\epsilon_q+\tilde\epsilon_k)t}
    \nonumber\\&
    -e^{ij_0(q-p)+ij(k-q)-i(\tilde\epsilon_q-\tilde\epsilon_p)t}
    \nonumber\\&
    -e^{ij_0(q-p))+ij(k-q)-i(\tilde\epsilon_q+\tilde\epsilon_p)t}
    \bigg)\,,
\end{align}
and
\begin{align}
    A^{ab}_{k,p}(t)=&i\sin\alpha(\cos\alpha-1)\nonumber\\&\times\frac{1}{N}\sum_{q}\bigg(
    e^{ij_0(k-q)+ij(q-p)+i(\tilde\epsilon_q+\tilde\epsilon_k)t}
    \nonumber\\&
    +e^{ij_0(k-q)+ij(q-p)-i(\tilde\epsilon_q-\tilde\epsilon_k)t}
    \nonumber\\&
    -e^{ij_0(q-p)+ij(k-q)+i(\tilde\epsilon_q-\tilde\epsilon_p)t}
    \nonumber\\&
    -e^{ij_0(q-p)+ij(k-q)+i(\tilde\epsilon_q+\tilde\epsilon_p)t}
    \bigg)\,.
\end{align}
When we take the average over the ground state of the commutator squared we find
\begin{align}
    C_{j,j_0}(t)=\frac{1}{N^2}&\sum_{k,p,k',p'}\bigg[B_{k,p}(t)B^*_{k',p'}(t)\left\langle b^\dagger_p b_kb^\dagger_{k'} b_{p'}\right\rangle
    \nonumber\\&
    +A^{ba}_{k,p}(t)[A^{ba}_{k',p'}(t)]^*\left\langle b^\dagger_p a_ka^\dagger_{k'} b_{p'}\right\rangle
    \bigg]\,.
\end{align}
This can now be calculated for particular examples of Hamiltonians and ground states.

For the Hamiltonian of the chiral modes we want a right moving linear dispersion. The simplest way to achieve this is to take $\mathcal{H}=-iv_F\tau^x\partial_x$ and we then have $\tilde\epsilon_k=-v_Fk$. Note here $\theta_k=0$. We then find
\begin{equation}
       \left|B_{k,p}(t)\right|^2=4\delta_{j-j_0,v_Ft}\sin^4\alpha\sin^2[(j-j_0)(k+p)]
\end{equation}
and 
\begin{equation}
       B_{k,k}(t)=2i\delta_{j-j_0,v_Ft}\sin^2\alpha\sin2(j_0-j)k\,.
\end{equation}
The next term is
\begin{align}
    \left|A^{ba}_{k,p}(t)\right|^2=4\delta_{j-j_0,v_Ft}\sin^2\alpha(\cos\alpha-1)^2&\cos^2(j-j_0)k\nonumber\\+
    4\delta_{j-j_0,-v_Ft}\sin^2\alpha(\cos\alpha-1)^2&\cos^2(j-j_0)p\,.
\end{align}
This term is an artifact of the empty band of left moving states, which we will neglect in the final analysis. The perturbation causes some of these states to become occupied during the dynamics and they are also involved in the scrambling.


%

\end{document}